\documentclass[reprint,superscriptaddress,nofootinbib,twocolumn,amsmath,amssymb,aps,prl]{revtex4-2}
\usepackage{graphicx}
\usepackage{braket}
\usepackage[colorlinks,citecolor=red,urlcolor=red,bookmarks=false,hypertexnames=true]{hyperref}
\newcommand{\cor}[1]{\left[ #1 \right]}
\newcommand{\abs}[1]{\left\vert #1 \right\vert}

\newcommand{\pare}[1]{\left( #1 \right)}

\begin{document}
\title{Smart Quantum Statistical Imaging beyond the Abbe-Rayleigh Criterion}
\author{Narayan Bhusal}
\thanks{These authors contributed equally.}
\affiliation{Quantum Photonics Laboratory, Department of Physics \& Astronomy, Louisiana State University, Baton Rouge, LA 70803, USA}
\author{Mingyuan Hong}
\thanks{These authors contributed equally.}
\affiliation{Quantum Photonics Laboratory, Department of Physics \& Astronomy, Louisiana State University, Baton Rouge, LA 70803, USA}
\author{Nathaniel R. Miller}
\thanks{These authors contributed equally.}
\affiliation{Quantum Photonics Laboratory, Department of Physics \& Astronomy, Louisiana State University, Baton Rouge, LA 70803, USA}
\author{Mario A Quiroz-Ju\'arez}
\affiliation{Departamento de Física, Universidad Autónoma Metropolitana Unidad Iztapalapa, San Rafael Atlixco 186, 09340 Ciudad México, M\'exico}
\author{Roberto de J. Le\'on-Montiel}
\affiliation{Instituto de Ciencias Nucleares, Universidad Nacional Aut\'onoma de M\'exico, Apartado Postal 70-543, 04510 Cd. Mx., M\'exico}
\author{Chenglong You}
\email{cyou2@lsu.edu}
\affiliation{Quantum Photonics Laboratory, Department of Physics \& Astronomy, Louisiana State University, Baton Rouge, LA 70803, USA}
\author{Omar S. Maga\~na-Loaiza}
\affiliation{Quantum Photonics Laboratory, Department of Physics \& Astronomy, Louisiana State University, Baton Rouge, LA 70803, USA}

\date{\today}
\begin{abstract}

The manifestation of the wave nature of light through diffraction imposes limits on the resolution of optical imaging. For over a century, the Abbe-Rayleigh criterion has been utilized to assess the spatial resolution limits of optical instruments. Recently, there has been an enormous impetus in overcoming the Abbe-Rayleigh resolution limit by projecting target light beams onto spatial modes. These conventional schemes for superresolution rely on a series of spatial projective measurements to pick up phase information that is used to boost the spatial resolution of optical systems. Unfortunately, these schemes require \textit{a priori} information regarding the coherence properties of ``unknown" light beams. Furthermore, they require stringent alignment and centering conditions that cannot be achieved in realistic scenarios. Here, we introduce a smart quantum camera for superresolving imaging. This camera exploits the self-learning features of artificial intelligence to identify the statistical fluctuations of unknown mixtures of light sources at each pixel. This is achieved through a universal quantum model that enables the design of artificial neural networks for the identification of quantum photon fluctuations.  Our camera overcomes the inherent limitations of existing superresolution schemes based on spatial mode projection. Thus, our work provides a new perspective in the field of imaging with important implications for microscopy, remote sensing, and astronomy. 


\end{abstract}

\maketitle

The spatial resolution of optical imaging systems is established by the diffraction of photons and the noise associated with their quantum fluctuations \cite{abbe1873beitrage, rayleigh1879xxxi, born2013principles, goodman2005introduction, magana2019quantum}. For over a century, the Abbe-Rayleigh criterion has been used to assess the diffraction-limited resolution of optical instruments \cite{born2013principles, won2009eyes}. At a more fundamental level, the ultimate resolution of optical instruments is established by the laws of quantum physics through the Heisenberg uncertainty principle \cite{stelzer2002beyond, kolobov2000quantum, stelzer2000uncertainty}. In classical optics, the Abbe-Rayleigh resolution criterion stipulates that an imaging system cannot resolve spatial features smaller than $\lambda/2\text{NA}$. In this case, $\lambda$ represents the wavelength of the illumination field, and $\text{NA}$ describes numerical aperture of the optical instrument \cite{abbe1873beitrage, rayleigh1879xxxi, born2013principles, editorial2009}.  Given the implications that overcoming the Abbe-Rayleigh resolution limit has for multiple applications, such as, microscopy, remote sensing, and astronomy \cite{pirandola2018advances, hell20152015, editorial2009, born2013principles}, there has been an enormous interest in improving the spatial resolution of optical systems \cite{tsang2009quantum, tsang2016quantum, hell1994breaking}. So far, optical superresolution has been achieved through spatial decomposition of eigenmodes \cite{paur2018tempering, tsang2016quantum, tamburini2006overcoming}. These conventional schemes rely on spatial projective measurements to pick up phase information that is used to boost spatial resolution of optical instruments \cite{tsang2016quantum,Steinberg17PRL,zhou2019quantum,Treps20Optica,Saleh18Optica,Liang21Optica}.

For almost a century, the importance of phase over amplitude information has constituted established knowledge for optical engineers \cite{goodman2005introduction, born2013principles,magana2019quantum}. Recently, this idea has been extensively investigated in the context of quantum metrology \cite{boto2000quantum, tang2016fault, parniak2018beating,you2020multiphoton, magana2019quantum}.  More specifically, it has been demonstrated that phase information can be used to surpass the Abbe-Rayleigh resolution limit for the spatial identification of light sources \cite{tsang2009quantum, giovannetti2009sub, Steinberg17PRL, zhou2019quantum, Treps20Optica}. For example, phase information can be obtained through mode decomposition by using projective measurements or demultiplexing of spatial modes \cite{tsang2016quantum, tamburini2006overcoming, Steinberg17PRL, zhou2019quantum, Treps20Optica}. Naturally, these approaches require \textit{a priori} information regarding the coherence properties of the, in principle, “unknown” light sources \cite{hell1994breaking, Saleh18Optica, Liang21Optica, tsang2016quantum}. Furthermore, these techniques impose stringent requirements on the alignment and centering conditions of imaging systems \cite{tamburini2006overcoming, Steinberg17PRL, zhou2019quantum, Treps20Optica, hell1994breaking, Saleh18Optica, OmarSciAdv2016, YangLight2017, Liang21Optica, tsang2016quantum}. Despite these limitations, most, if not all, the current experimental protocols have relied on spatial projections and demultiplexing in the Hermite-Gaussian, Laguerre-Gaussian, and parity basis \cite{tamburini2006overcoming, zhou2019quantum, Steinberg17PRL, Saleh18Optica, zhou2019quantum, Treps20Optica, Liang21Optica, tsang2016quantum}.

\begin{figure*}[!t]
  \centering
 \includegraphics[width=1\textwidth]{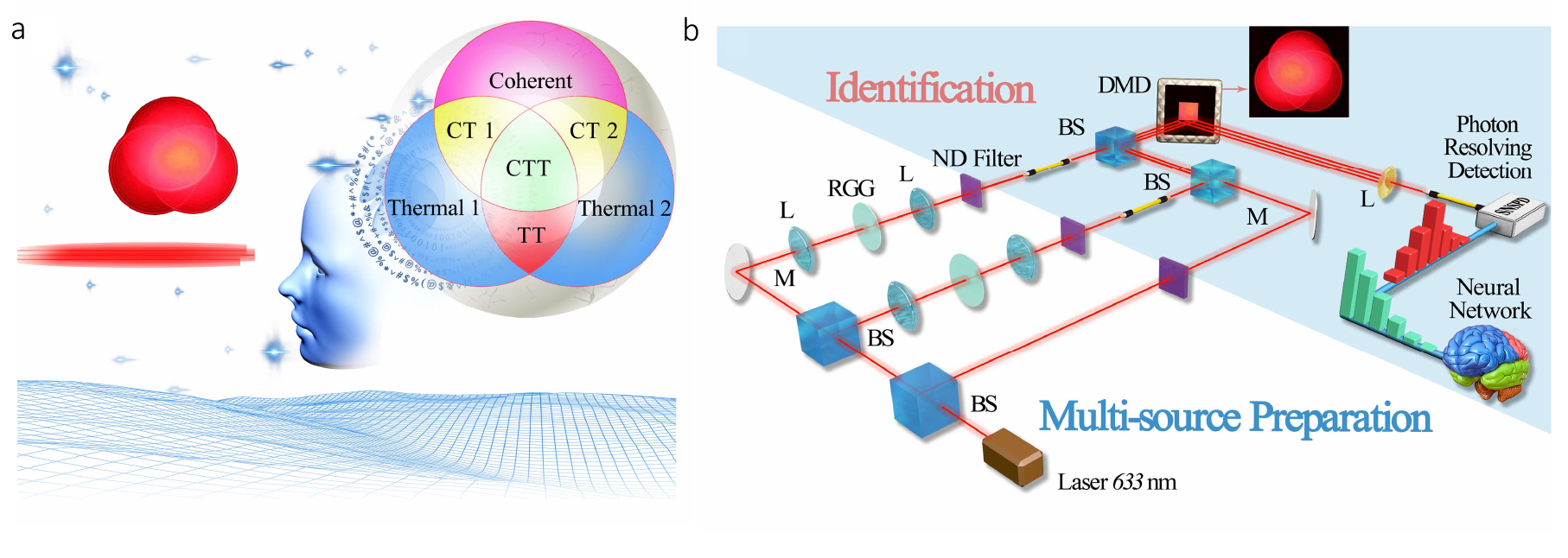}
\caption{Conceptual illustration and schematic of our experimental setup to demonstrate superresolving imaging. 
The illustration in \textbf{a} depicts a scenario where diffraction limits the resolution of an optical instrument for remote imaging. In our protocol, an artificial neural network enables the identification of the photon statistics that characterize the point sources that constitute a target object. In this case, the point sources emit either coherent or thermal photons. Remarkably, the neural network is capable of identifying the corresponding photon fluctuations and their combinations, for example coherent-thermal (CT1, CT2), thermal-thermal (TT) and coherent-thermal-thermal (CTT). This capability allows us to boost the spatial resolution of optical instruments beyond the Abbe-Rayleigh resolution limit.
The experimental setup in \textbf{b} is designed to generate two independent thermal and one coherent light sources.  The three sources are produced from a continuous-wave (CW) laser at $633 \text{ nm}$. The CW laser beam is divided by two beam splitters (BS) to generate three spatial modes, two of which are then passed through rotating ground glass (RGG) disks to produce two independent thermal light beams. The three light sources, with different photon statistics, are attenuated using neutral density (ND) filters and then combined to mimic a remote object such as the one shown in the inset of \textbf{b}. This setup enables us to generate multiple sources with tunable statistical properties.  The generated target beam is then imaged onto a digital micro-mirror device (DMD) that we use to perform raster scanning. The photons reflected off the DMD are collected and measured by a single-photon detector. Our protocol is formalized by performing photon-number-resolving detection \cite{you2020identification}. The characteristic quantum fluctuations of each light source are identified by an artificial neural network. This information is then used to produce a high-resolution image of the object beyond the diffraction limit. }
\label{schematic}
\end{figure*}

The quantum statistical fluctuations of photons  establish the nature of light sources \cite{You2021naturecomm,mandel78,magana2019multiphoton,you2020identification, gerry2005introductory}. As such, these fundamental properties are not affected by the spatial resolution of an optical instrument \cite{gerry2005introductory}. Here,  we demonstrate that measurements of the quantum statistical properties of a light field enable imaging beyond the Abbe-Rayleigh resolution limit. This is performed by exploiting the self-learning features of artificial intelligence to identify the statistical fluctuations of photon mixtures \cite{you2020identification}. More specifically, we demonstrate a smart quantum camera with the capability to identify photon statistics at each pixel. For this purpose, we introduce a universal quantum model that describes the photon statistics produced by the scattering of an arbitrary number of light sources. This model is used to design and train artificial neural networks for the identification of light sources. Remarkably, our scheme enables us to overcome inherent limitations of existing superresolution protocols based on spatial mode projections and multiplexing \cite{tsang2016quantum, tamburini2006overcoming, zhou2019quantum, Steinberg17PRL, Saleh18Optica, zhou2019quantum, Treps20Optica, Liang21Optica}.

The conceptual schematic behind our experiment is depicted in 
Fig. \ref{schematic}\textbf{a}. This camera utilizes an artificial neural network to identify the photon statistics of each point source that constitutes a target object. The description of the photon statistics produced by the scattering of an arbitrary number of light sources is achieved through a general model that relies on the quantum theory of optical coherence introduced by Sudarshan and Glauber \cite{sudarshan1963equivalence, glauber1963quantum, gerry2005introductory}. We use this model to
design and train a neural network capable of identifying light sources at each pixel of our camera. This unique feature is achieved by performing photon-number-resolving detection \cite{you2020identification}. The sensitivity of this camera is limited by the photon fluctuations, as stipulated by the Heisenberg uncertainty principle, and not by the Abbe-Rayleigh resolution limit \cite{gerry2005introductory, magana2019quantum}. 

\begin{figure*}[!ht]
\centering
\includegraphics[width=1\textwidth]{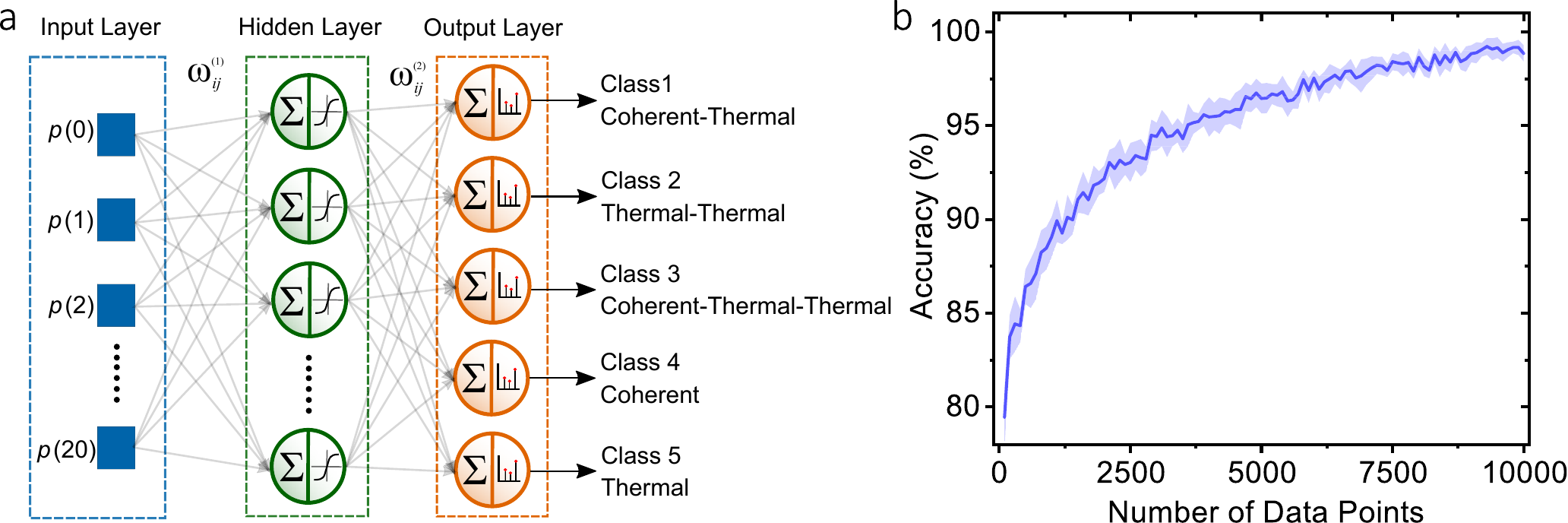}
\caption{The scheme of the two-layer neural network used to identify the photon statistics produced by a combination of three sources is shown in \textbf{a}. The computational model consists of an input layer, a hidden layer of sigmoid neurons, and a Softmax output layer. The training of our neural network through Eqs. \ref{pthcoh} and \ref{pthcohdis} enables the efficient identification of five classes of photon statistics. Each class is characterized by a $g^{(2)}$ function, which is defined by a specific combination of light sources \cite{you2020identification}. In our experiment, these classes correspond to the characteristic photon statistics produced by coherent or thermal light sources and their combinations. For example, coherent-thermal, thermal-thermal, or coherent-thermal-thermal.  The figure in \textbf{b} shows the performance of our neural network as a function of the number of data samples used each time in the testing process. The classification accuracy for the five possible complex classes of light is 80\% with 100 data points. Remarkably, the performance of the neural network increases to approximately 95\% when we use 3500 data points in each test sample.}
\label{nnAccuracy}
\end{figure*}

In general, realistic imaging instruments deal with the detection of multiple light sources. These sources can be either distinguishable or indistinguishable \cite{born2013principles, gerry2005introductory}. The combination of indistinguishable sources can be represented by either coherent or incoherent superpositions of light sources characterized by Poissonian (coherent) or super-Poissonionan (thermal) statistics \cite{gerry2005introductory}. In our model, we first consider the indistinguishable detection of $N$ coherent and $M$ thermal sources. For this purpose, we make use of the P-function  $P_{\text{coh}}(\gamma)=\delta^2 (\gamma-\alpha_{k})$ to model the contributions from the $k$th coherent source with the corresponding complex amplitude $\alpha_k$ \cite{sudarshan1963equivalence, glauber1963quantum}. The total complex amplitude associated to the superposition of an arbitrary number of light sources is given by $\alpha_{\text{tot}}=\sum_{k=1}^{N}\alpha_k$.
In addition, the P-function for the $l$th thermal source, with the corresponding mean photon numbers $\bar{m}_l$, is defined as $P_{\text{th}}(\gamma)=(\pi\bar{m}_l)^{-1}\exp{(-|\gamma|^2}/\bar{m}_l)$. The total number of photons attributed to the $M$ number of thermal sources is defined as $m_{\text{tot}}=\sum_{l=1}^{M}\bar{m}_l$. 
These quantities allow us to calculate the P-function for the multisource system as
\textcolor{black}{
\begin{equation}
\label{pfun}
\begin{aligned}
P_{\text{th-coh}}(\gamma)& = \int\cdots\int  P_{N+M}(\gamma-\gamma_{N+M-1})\\
&  \times \cor{\prod_{i=2}^{N+M-1} P_i(\gamma_i-\gamma_{i-1})d^{2}\gamma_i} P_{1}(\gamma_{1})d^{2}\gamma_{1}.
\end{aligned}
\end{equation}}
This approach enables the analytical description of the photon-number distribution $p_{\text{th-coh}}(n)$ associated to the detection of an arbitrary number of indistinguishable light sources. This is calculated as $p_{\text{th-coh}}(n)=\bra{n} \hat{\rho}_{\text{th-coh}} \ket{n}$, where $\rho_{\text{th-coh}}=\int{P_{\text{th-coh}}(\gamma) \ket{\gamma}\bra{\gamma}}d^2\gamma$. After algebraic manipulation (see Supplementary Information), we obtain the following photon-number distribution
\begin{equation}
\label{pthcoh}
\begin{aligned}
p_{\text{th-coh}}(n)=&\frac{\left(m_{\text{tot}}\right)^{n} \exp \left(-\left(|\alpha_\text{tot}|\right)^{2} / m_{\text {tot}}\right)}{\pi\left(m_{\text{tot}}+1\right)^{n+1}}\\
& \times \sum_{k=0}^{n} \frac{1}{k !(n-k) !} \Gamma\left(\frac{1}{2}+n-k\right) \Gamma\left(\frac{1}{2}+k\right)\\
& { }_{1} F_{1}\left(\frac{1}{2}+n-k; \frac{1}{2}; \frac{(\operatorname{Re}[\alpha_\text{tot}])^{2}}{m_{\text{tot}}\left(m_{\text{tot}}+1\right)}\right)\\
& { }_{1} F_{1}\left(\frac{1}{2}+k;\frac{1}{2}; \frac{(\operatorname{Im}[\alpha_\text{tot}])^{2}}{m_{\text{tot}}\left(m_{\text{tot}}+1\right)}\right),
\end{aligned}
\end{equation}
\\
\\
where $\Gamma(z)$ and ${ }_{1} F_{1}(a;b;z)$ are the Euler gamma and the Kummer confluent hypergeometric functions, respectively. This probability function enables the general description of the photon statistics produced by any indistinguishable combination of light sources. Thus, the photon distribution produced by the distinguishable detection of $N$ light sources can be simply obtained by performing a discrete convolution of Eq. \ref{pthcoh} as
\begin{equation}
\label{pthcohdis}
\begin{aligned}
p_{\text{tot}}(n)=&\sum_{{m_1}=0}^{n} \sum_{{m_2}=0}^{n-m_1}  \dotsb \sum_{{m_{N-1}}=0}^{n-\sum^{N-1}_{j=1}m_j} p_1(m_1)p_2(m_2)\dotsb \\
& p_{N-1}(m_{N-1}) p_N(n-\sum^{N-1}_{j=1}m_j).
\end{aligned}
\end{equation}
The combination of Eq. \ref{pthcoh} and Eq. \ref{pthcohdis} allows the classification of photon-number distributions for any combination of light sources. 

We demonstrate our proof-of-principle quantum camera using the experimental setup shown in Fig. \ref{schematic}\textbf{b}. For this purpose, we use a continuous-wave laser at $ 633 \text{nm}$ to produce either coherent, or incoherent superpositions of distinguishable, indistinguishable, or partially distinguishable light sources.  In this case, the combination of photon sources, with tunable statistical fluctuations, acts as our target object. Then, we image our target object onto a digital micro-mirror device (DMD) that is used to implement raster scanning. This is implemented by selectively turning on and off groups of pixels in our DMD. The light reflected off the DMD is measured by a single-photon detector that allows us to perform photon-number-resolving detection. This is implemented through the technique described in ref. \cite{you2020identification}.

\begin{figure}[t!]
\centering
\includegraphics[width=\linewidth]{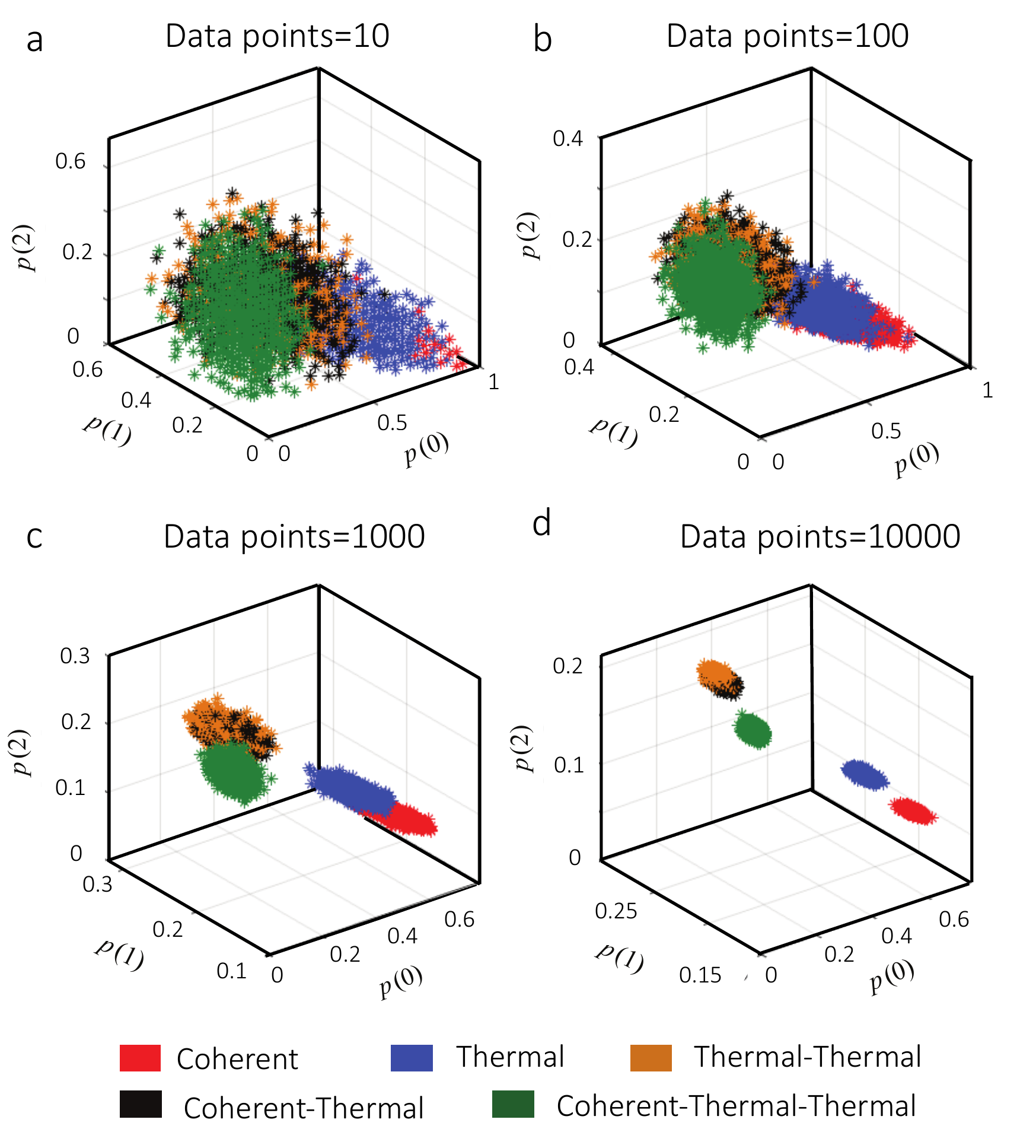}
\caption{Projection of the feature space on the plane defined by the probabilities $p(0)$, $p(1)$, and $p(2)$. The red points correspond to the photon statistics for coherent light, and the blue points indicate the photon statistics for thermal light fields. Furthermore, the brown dots represent the photon statistics produced by the scattering of two thermal light sources, and the black points show the photon statistics for a mixture of photons emitted by one coherent and one thermal source. The corresponding statistics for a mixture of one coherent and two thermal sources are indicated in green. As shown in \textbf{a}, the distributions associated to the multiple sources obtained for 10 data points are confined to a small region of the feature space. A similar situation prevails in \textbf{b} for 100 data points. As shown in panel \textbf{c}, the distributions produced with 1000 data points occupy different regions, although brown and black points keep closely intertwined. Finally, the separated distributions obtained with 10000 data points in \textbf{d} enable efficient identification of light sources.}
\label{fig:numberdatapoint}
\end{figure}

\begin{figure*}[!htbp]
  \centering
 \includegraphics[width=0.95\textwidth]{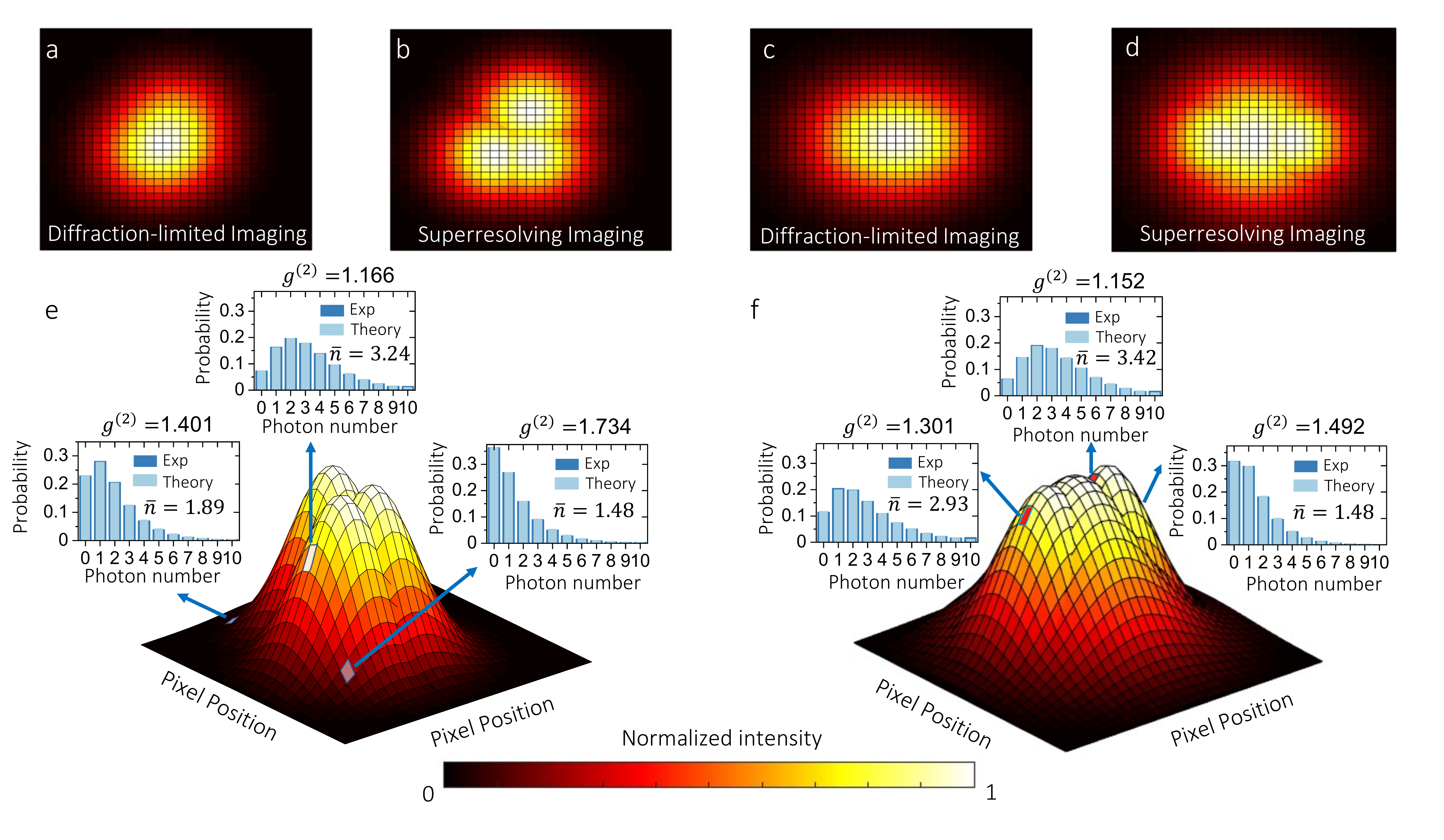}
\caption{Experimental superresolving imaging. The plot in \textbf{a} shows the combined intensity profile of the three partially distinguishable sources. As stipulated by the Abbe-Rayleigh resolution criterion, the transverse separations among the sources forbid their identification. As shown in \textbf{b}, our smart quantum camera enables superresolving imaging of the remote sources.  In \textbf{c} and \textbf{d}, we show another experimental realization of our protocol for a different distribution of light sources. In this case, two small sources are located inside the point-spread function of a third light source. The figures in \textbf{e} and \textbf{f} correspond to the inferred spatial distributions based on the experimental pixel-by-pixel imaging used to produce \textbf{b} and \textbf{d}. The insets in \textbf{e} and \textbf{f} show photon-number probability distributions for three pixels, the theory bars were obtained through Eqs. \ref{pthcoh} and \ref{pthcohdis}. These results demonstrate the potential of our technique to outperform conventional diffraction-limited imaging.}
\label{profiles}
\end{figure*}

The equations above allow us to implement a multi-layer feed-forward network for the identification of the quantum photon fluctuations of the point sources of a target object. The structure of the network consists of a group of interconnected neurons arranged in layers. Here, the information flows only in one direction, from input to output \cite{svozil1997introduction, bhusal2021spatial}. As indicated in Fig. \ref{nnAccuracy}\textbf{a}, our network comprises two layers, with ten sigmoid neurons in the hidden layer (green neurons) and five softmax neurons in the output layer (orange neurons). In this case, the input features represent the probabilities of detecting $n$ photons at a specific pixel, $p(n)$, whereas the neurons in the last layer correspond to the classes to be identified. The input vector is then defined by twenty-one features corresponding to $n$=0,1,...,20. In our experiment, we define five classes that we label as: coherent-thermal (CT), thermal-thermal (TT), coherent-thermal-thermal (CTT), coherent (C), and thermal (T). If the brightness of the experiment remains constant, these classes can be directly defined through the photon-number distribution described by Eqs. \ref{pthcoh} and \ref{pthcohdis}. However, if the brightness of the sources is modified, the classes can be defined through the $g^{(2)}=1+\left(\left\langle(\Delta \hat{n})^{2}\right\rangle-\langle\hat{n}\rangle\right) /\langle\hat{n}\rangle^{2}$, which is intensity-independent \cite{you2020identification, You2021naturecomm}. The parameters in the $g^{(2)}$ function can also be calculated from Eqs. \ref{pthcoh} and \ref{pthcohdis}. It is important to mention that the output neurons provide a probability distribution over the predicted classes \cite{Goodfellow2016,bishop2006pattern}. The training details of our neural networks can be found in the Methods section. 


We test the performance of our neural network through the classification of a complex mixture of photons produced by the combination of one coherent with two thermal light sources. The accuracy of our trained neural network is reported in Fig. \ref{nnAccuracy}\textbf{b}. In our setup, the three partially overlapping sources form five classes of light with different mean photon numbers and photon statistics. We exploit the functionality of our artificial neural network to identify the underlying quantum fluctuations that characterize each kind of light. We calculate the accuracy as the ratio of true positive and true negative to the total of input samples during the testing phase. Fig. \ref{nnAccuracy}\textbf{b} shows the overall accuracy as a function of the number of data points used to build the probability distributions for the identification of the multiple light sources using a supervised neural network. The classification accuracy for the mixture of three light sources is 80\% with 100 photon-number-resolving measurements. The performance of the neural networks increases to approximately 95\% when we use 3500 data points to generate probability distributions.

The performance of our protocol for light identification can be understood through the distribution of light sources in the probability space shown in Fig. \ref{fig:numberdatapoint}. Here we show the projection of the feature space on the plane defined by the probabilities $p(0)$, $p(1)$, and $p(2)$ for different number of data points. Each point is obtained from an experimental probability distribution. As illustrated in Fig. \ref{fig:numberdatapoint}\textbf{a}, the distributions associated to the multiple sources obtained for 10 data points are confined to a small region of the feature space. This condition makes extremely hard the identification of light sources with 10 sets of measurements. A similar situation can be observed for the distribution in Fig. \ref{fig:numberdatapoint}\textbf{b} that was generated using 100 data points. As shown in panel Fig. \ref{fig:numberdatapoint}\textbf{c}, the separations in the distributions produced with 1000 data points occupy different regions, although brown and black points keep closely intertwined. These conditions enable one to identify multiple light sources. Finally, the separated distributions obtained with 10000 data points in Fig. \ref{fig:numberdatapoint}\textbf{d} enable efficient identification of light sources. These probability space diagrams explain the performances reported in Fig. \ref{nnAccuracy}. An interesting feature of Fig. \ref{fig:numberdatapoint} is the fact that the distributions in the probability space are linearly separable.

\begin{figure*}[t!]
 \centering
 \includegraphics[width=0.95\textwidth]{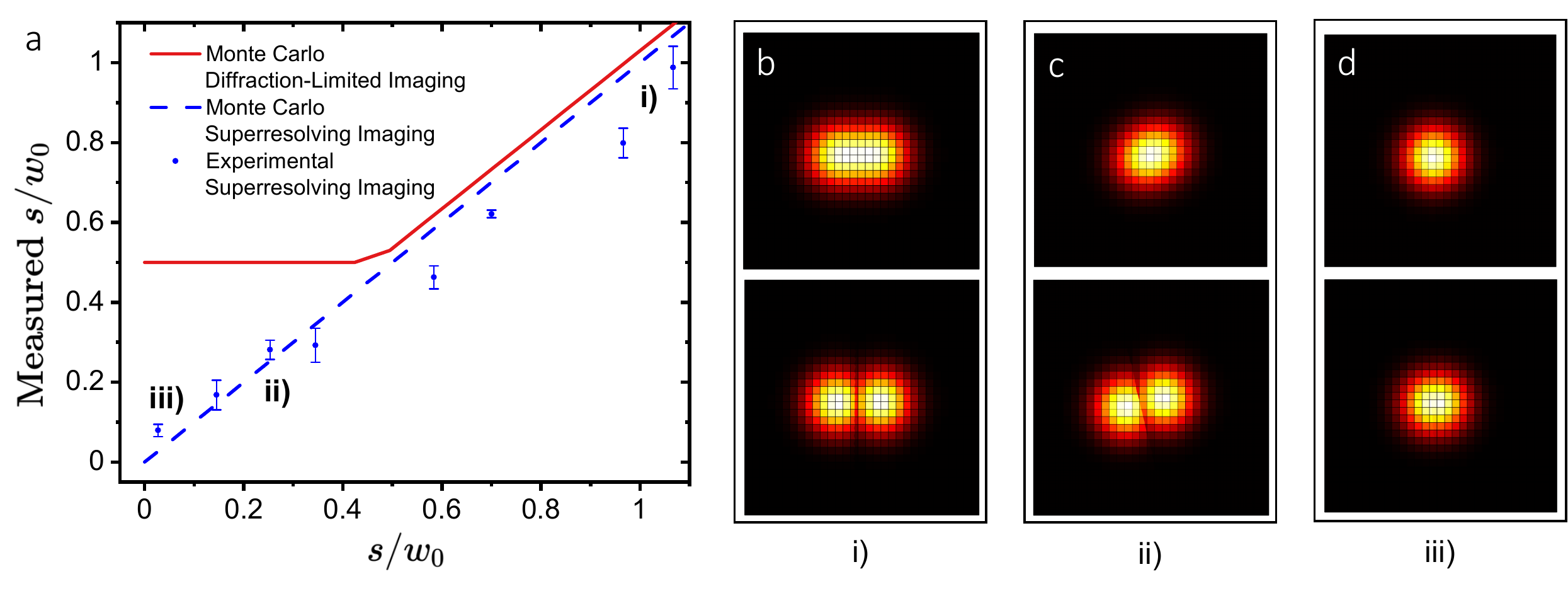}
 \caption{Comparison between the spatial resolution of our camera and direct imaging. Here the distance is normalized by the beam radius for easy identification of the Abbe-Rayleigh limit. As shown in \textbf{a}, the red line is the result of a Monte-Carlo simulation for traditional intensity based direct imaging.  The plateau is the area where the algorithm becomes unstable. The dotted blue line represents the limit for our supperresolving imaging method, where perfect classification of each pixel is assumed. The blue dots represent the experimental data collected with our camera for superresolving imaging. The experimental points demonstrate the potential of our technique for identifying spatial features beyond the Abbe-Rayleigh resolution criterion. The 
 first row in the panels from \textbf{b} to \textbf{d} shows the reconstructed spatial profiles 
 obtained through direct imaging whereas the second row shows the superresolving images obtained with our technique. The panel in \textbf{b} shows the spatial profiles for the experimental point $i)$. This corresponds to the experimental detection of two sources with the largest separation. The spatial profiles in \textbf{c} correspond to the experimental point labeled as $ii)$. Finally, the panel in \textbf{d} shows the spatial distributions for the experimental point with the smallest separation, this is labeled as $iii)$.}
 \label{distances}
\end{figure*}

As demonstrated in Fig. \ref{profiles}, the identification of the quantum photon fluctuations at each pixel of our camera enables us to demonstrate superresolving imaging.  In our experiment we prepared each source to have a mean photon number between 1 and 1.5 for the brightest pixel. The raster-scan image of a target object composed of multiple partially distinguishable sources in Fig. \ref{profiles}\textbf{a} illustrates the performance of conventional imaging protocols limited by diffraction \cite{goodman2005introduction, won2009eyes, stelzer2002beyond, kolobov2000quantum}. In this case, it is practically impossible to identify the multiple sources that constitute the target object. Remarkably, as shown in Fig. \ref{profiles}\textbf{b}, our protocol provides a dramatic improvement of the spatial resolution of the imaging system. In this case, it becomes clear the presence of the three emitters that form the remote object. The estimation of separations among light sources is estimated through a fit over the classified pixel-by-pixel image. Additional details can be found in the Methods section. In Figs. \ref{profiles}\textbf{c} and \textbf{d}, we demonstrate the robustness of our protocol by performing superresolving imaging for a different configuration of light sources. In this case, two small sources are located inside the point-spread function of a third light source. As shown in Fig. \ref{profiles}\textbf{c}, the Abbe-Rayleigh limit forbids the identification of light sources. However, we demonstrate substantial improvement of spatial resolution in Fig. \ref{profiles}\textbf{d}. The plots in Figs. \ref{profiles}\textbf{e} and \textbf{f} correspond to the inferred spatial distributions based on the experimental pixel-by-pixel imaging used to produce Figs. \ref{profiles}\textbf{b} and \textbf{d}. The insets in Figs. \ref{profiles}\textbf{e} and \textbf{f} show photon-number probability distributions for three pixels. The theoretical photon-number distributions in Fig.\ref{profiles}\textbf{e} and \textbf{f} are obtained through a procedure of least square regression \cite{massaron2016regression}. Here the least squares difference between the measured and theoretical probability distribution was minimized for $0\leq n\leq 6$. The sources were assumed to be partially distinguishable  allowing the theoretical distribution to be defined by Eqs. \ref{pthcoh} and Eq. \ref{pthcohdis}.  The combined mean photon numbers of each source generated for the fit totals the measured mean photon number (see Methods section). Our scheme enables the use of the photon-number distributions or their corresponding $g^{(2)}$ to characterize light sources. This allows us to determine each pixel's corresponding statistics, regardless of the mean photon numbers of the sources in the detected field \cite{you2020identification, You2021naturecomm}.


We now provide a quantitative characterization of our superresolving imaging scheme based on the identification of photon statistics. We demonstrate that our smart camera for superresolving imaging can capture small spatial features that surpass the resolution capabilities of conventional schemes for direct imaging \cite{abbe1873beitrage, rayleigh1879xxxi, born2013principles, goodman2005introduction, magana2019quantum}. Consequently, as shown in Fig. \ref{distances}, our camera enables the possibility of performing imaging beyond the Abbe-Rayleigh criterion. In this case, we performed multiple experiments in which a superposition of partially distinguishable sources were imaged. The superposition was prepared using one coherent and one thermal light source. In Fig. \ref{distances}\textbf{a}, we plot the predicted transverse separation $s$ normalized by the Gaussian beam waist radius $w_0$ for both protocols. Here $w_0=\lambda/\pi\text{NA}$, this parameter is directly obtained from our experiment.   As demonstrated in Fig. \ref{distances}\textbf{a}, our protocol enables one to resolve spatial features for sources with small separations even for diffraction-limited conditions. As expected for larger separation distances, the performance of our protocol matches the accuracy of intensity measurements. This is further demonstrated by the spatial profiles shown from Fig. \ref{distances}\textbf{b} to \textbf{d}. The first row shows spatial profiles for three experimental points in Fig. \ref{distances}\textbf{a} obtained through direct imaging whereas the images in the second row were obtained using our scheme for superresolving imaging. The spatial profiles in Fig. \ref{distances}\textbf{b} show that both imaging techniques lead to comparable resolutions and the correct identification of the centroids of the two sources. However, as shown in Fig. \ref{distances}\textbf{c} and \textbf{d}, our camera outperforms direct imaging when the separations decrease. Here, the actual separation is smaller than $w_0/2$ for both cases. It is worth noticing that in this case, direct imaging cannot resolve spatial features of the sources. Here, the predictions of direct imaging become unstable and erratic. Remarkably, our simulations show an excellent agreement with the experimental data obtained for our scheme for superresolving imaging (see Methods section).

In conclusion, we demonstrated a robust quantum camera that enables superresolving imaging beyond the Abbe-Rayleigh resolution limit. Our scheme for quantum statistical imaging exploits the self-learning features of artificial intelligence to identify the statistical fluctuations of truly unknown mixtures of light sources. This particular feature of our scheme relies on a universal model based on the theory of quantum coherence to describe the photon statistics produced by the scattering of an arbitrary number of light sources. We demonstrated that the measurement of the quantum statistical fluctuations of photons enables one to overcome inherent limitations of existing superresolution protocols based on spatial mode projections \cite{tsang2016quantum,Steinberg17PRL,zhou2019quantum,Treps20Optica,Saleh18Optica,Liang21Optica}. We believe that our work represents a new paradigm in the field of optical imaging with important implications for microscopy, remote sensing, and astronomy \cite{magana2019quantum, won2009eyes, stelzer2002beyond, kolobov2000quantum, stelzer2000uncertainty,editorial2009, pirandola2018advances}.

\section*{Acknowledgments}
N.B., M.H., C.Y., and O.S.M.L. acknowledge support from the Army Research Office (ARO) under the grant no. W911NF-20-1-0194, the U.S. Department of Energy, Office of Basic Energy Sciences, Division of Materials Sciences and Engineering under Award DE-SC0021069, and the Louisiana State University (LSU) Board of Supervisors for the LSU Leveraging Innovation for Technology Transfer (LIFT2) Grant under the grant no. LSU-2021-LIFT-004. N.M. thanks support from Department of Physics \& Astronomy of Louisiana State University. R.J.L.M. thankfully acknowledges financial support by CONACyT under the project CB-2016-01/284372, and by DGAPA-UNAM, under the project UNAM-PAPIIT IN102920.

\section{COMPETING INTERESTS}
The authors declare no competing interests.

\section{DATA AVAILABILITY}
The data sets generated and/or analyzed during this study are available from the corresponding author or last author on reasonable request.

\section*{Methods}
\subsection{Training of NN}
For the sake of simplicity, we split the functionality of our neural network  into two phases: the training and testing phase. In the first phase, the training data is fed to the network multiple times to optimize the synaptic weights through a scaled conjugate gradient back-propagation algorithm \cite{moller1993scaled}. This optimization seeks to minimize the Kullback-Leibler divergence distance between predicted and the real target classes \cite{kullback1951information, kullback1997information}. At this point, the training is stopped if the loss function does not decrease within 1000 epochs \cite{prechelt1998early}. In the test phase, we assess the performance of the algorithm by introducing an unknown set of data during the training process. For both phases,  we prepare a data-set consisting of one thousand experimental measurements of photon statistics for each of the five classes. This process is formalized by considering different numbers of data points: 100, 500, ..., 9500, 10000. Following a standardized ratio for statistical learning, we divide our data into training (70\%), validation (15\%), and testing (15\%) sets \cite{crowther2005method}.  The networks were trained using the neural network toolbox in MATLAB, which runs on a computer Intel Core i7–4710MQ CPU (@2.50GHz) with 32GB of RAM.

\subsection{Fittings}
To determine the optimal fits for Fig. \ref{profiles}\textbf{e} and \textbf{f} we design a search space based on Eqs. \ref{pthcoh} and \ref{pthcohdis}.  To do so we first found the mean photon number of the input pixel, which will later be applied to constrain the search space.  From here we allowed for the existence of up to three distinguishable modes which will be combined according to Eq. \ref{pthcohdis}.  Each of the modes contains an indistinguishable combination of up to one coherent and two thermal sources whose number distribution is given by Eq. \ref{pthcoh}.  The total combination results in partially distinguishable combination and provides the theoretical model for our experiment.  From here our search space is 

\begin{align*}
    \sqrt{\sum_{n=0}(p_{\text{exp}}(n)-p_{\text{th}}(n|\vec{n}_{1,t},\vec{n}_{2,t},\vec{n}_c))^2},
\end{align*}   
where $\vec{n}_{i,t}$ and $\vec{n}_{c}$ are the mean photon numbers of that each thermal or coherent source contributes to each distinguishable mode respectively.  The mean photon numbers of each source must add up to the experimental mean photon number, constraining the search.  A linear search was then performed over the predicted mean photon numbers and the minimum was returned, providing the optimal fit.

\subsection{Monte-Carlo Simulation of the Experiment}
To demonstrate a consistent improvement over traditional methods, we also simulated the experiment using two beams, a thermal and a coherent, with Gaussian point spread functions over a 128$\times$128 grid of pixels. At each pixel, the mean photon number for each source is provided by the Gaussian point spread function, which is then used to create the appropriate distinguishable probability distribution as given in Eq. \ref{pthcohdis}, creating a 128$\times$128 grid of photon number distributions. The associated class data for these distributions will then be fitted using to a set of pre-labeled disks using a genetic algorithm.  This recreates our method in the limits of perfect classification. Each of these distributions is then used to simulate photon-number resolving detection. This data is then used to create a normalized intensity for the classical fit. We fit the image to a combination of Gaussian PSFs. This process is repeated ten times for each separation in order to average out fluctuations in the fitting. When combining the results of the intensity fits they are first divided into two sets.  One set has the majority of fits return a single Gaussian, while the other returned two Gaussian the majority of the time. The set identified as only containing a single Gaussian is then set at the Abbe-Rayleigh diffraction limit, while the remaining data is used in a linear fit. This causes the sharp transition between the two sets of data.
\bibliography{main.bib}

\newpage
\onecolumngrid 
\section*{Supplementary Material: Derivation of the many-source photon-number distribution}

Let us start by considering the indistinguishable detection of $N$ coherent and $M$ thermal independent sources. To obtain the combined photon distribution, we make use of the Glauber-Sudarshan theory of coherence \cite{glauber1963quantum,sudarshan1963equivalence}. Thus, we start by writing the P-functions associated to the fields produced by the indistinguishable coherent and thermal sources, that is, we write
\begin{equation}\label{Eq:Pcoh}
P_{\text{coh}}\pare{\alpha} = \int P^{\text{coh}}_{N}\pare{\alpha - \alpha_{_{N-1}}}P^{\text{coh}}_{N-1}\pare{\alpha_{_{N-1}}-\alpha_{_{N-2}}}\cdots P^{\text{coh}}_{2}\pare{\alpha_{_{2}}-\alpha_{_{1}}}P^{\text{coh}}_{1}\pare{\alpha_{_{1}}}d^{2}\alpha_{_{N-1}}d^{2}\alpha_{_{N-2}}\cdots d^{2}\alpha_{_{2}}d^{2}\alpha_{_{1}},
\end{equation}
\begin{equation}\label{Eq:Pth}
P_{\text{th}}\pare{\alpha} = \int P^{\text{th}}_{M}\pare{\alpha - \alpha_{_{M-1}}}P^{\text{th}}_{M-1}\pare{\alpha_{_{M-1}}-\alpha_{_{M-2}}}\cdots P^{\text{th}}_{2}\pare{\alpha_{_{2}}-\alpha_{_{1}}}P^{\text{th}}_{1}\pare{\alpha_{_{1}}}d^{2}\alpha_{_{M-1}}d^{2}\alpha_{_{M-2}}\cdots d^{2}\alpha_{_{2}}d^{2}\alpha_{_{1}},
\end{equation}
with $P_{\text{coh}}\pare{\alpha}$ and $P_{\text{th}}\pare{\alpha}$ standing for the P-functions of the combined $N$-coherent and $M$-thermal sources, respectively. In both equations, $\alpha$ stands for the complex amplitude as defined for coherent states $\ket{\alpha}$, and the individual-source P-functions are defined as
\begin{equation}\label{Eq:P1coh}
P^{\text{coh}}_{k}\pare{\alpha} = \delta^{2}\pare{\alpha - \alpha_{k}},
\end{equation}
\begin{equation}\label{Eq:P1th}
P^{\text{th}}_{l}\pare{\alpha} = \frac{1}{\pi\bar{m}_{l}}\exp\pare{-\abs{\alpha}^{2}/\bar{m}_{l}},
\end{equation}
where $P^{\text{coh}}_{k}\pare{\alpha}$ corresponds to the P-function of $k$th coherent source, with mean photon number $\bar{n}_{k} = \abs{\alpha_{k}}^2$, and $P^{\text{th}}_{l}\pare{\alpha}$ describes the $l$th thermal source, with mean photon number $\bar{m}_{l}$.
\\
Now, by substituting Eq. (\ref{Eq:P1coh}) into Eq. (\ref{Eq:Pcoh}), and Eq. (\ref{Eq:P1th}) into Eq. (\ref{Eq:Pth}), we obtain
\begin{equation}
P_{\text{coh}}\pare{\alpha} = \delta^{2}\pare{\alpha - \sum_{k=1}^{N}\alpha_{k}},
\end{equation}
\begin{equation}
P_{\text{th}}\pare{\alpha} = \frac{1}{\pi\sum_{l=1}^{M}\bar{m}_{l}}\exp\pare{-\frac{\abs{\alpha}^{2}}{\sum_{l=1}^{M}\bar{m}_{l}}}.
\end{equation}
\\
We can finally combine the thermal and coherent sources by writing
\begin{equation}\label{Eq:Pth-coh}
P_{\text{th-coh}}\pare{\alpha} = \int P_{\text{th}}\pare{\alpha - \alpha'}P_{\text{coh}}\pare{\alpha'}d^2\alpha'.
\end{equation}
Note that this expression enables the analytical description for the photon-number distribution $p_{\text{th-coh}}\pare{n}$ of an arbitrary number of indistinguishable sources measured by a quantum detector. Also notice that Eq. (\ref{Eq:Pth-coh}) is equivalent to Eq. (1) in the main text. More specifically, we can write
\begin{equation}\label{Eq:photon_th_coh}
p_{\text{th-coh}}\pare{n} = \bra{n}\hat{\rho}_{\text{th-coh}}\ket{n},
\end{equation}  
where 
\begin{equation}\label{Eq:rho}
\hat{\rho}_{\text{th-coh}} = \int P_{\text{th-coh}}\pare{\alpha}\ket{\alpha}\bra{\alpha}d^{2}\alpha,
\end{equation}
describes the the density matrix of the quantum states of the combined thermal-coherent field at the quantum detector.

Thus, by substituting Eq. (\ref{Eq:Pth-coh}) into (\ref{Eq:rho}) and (\ref{Eq:photon_th_coh}), we find that the photon distribution of the combined fields is given by

\begin{equation}\label{pthcoh}
\begin{split}
p_{\text{th-coh}}(n)=&\frac{\left(m_{\text{tot}}\right)^{n} \exp \left(-\abs{\alpha_{\text{tot}}}^{2} / m_{\text{tot}}\right)}{\pi\left(m_{\text{tot}}+1\right)^{n+1}}\sum_{k=0}^{n} \frac{1}{k !(n-k) !} \Gamma\left(\frac{1}{2}+n-k\right) \Gamma\left(\frac{1}{2}+k\right)\\
& \times {}_1F_{1}\left(\frac{1}{2}+n-k; \frac{1}{2}; \frac{(\operatorname{Re}[\alpha_{\text{tot}}])^{2}}{m_{\text{tot}}\left(m_{\text{tot}}+1\right)}\right){}_1F_{1}\left(\frac{1}{2}+k;\frac{1}{2}; \frac{(\operatorname{Im}[\alpha_{\text{tot}}])^{2}}{m_{\text{tot}}\left(m_{\text{tot}}+1\right)}\right),
\end{split}
\end{equation}
\\
with $m_{\text{tot}} = \sum_{l=1}^{M}\bar{m}_{l}$ and $\alpha_{\text{tot}} = \sum_{k=1}^{N}\alpha_{k}$. In this final result, which corresponds to Eq. (2) of the main text, $\Gamma(z)$ and ${}_1F_1(a;b;z)$ are the Euler gamma and the Kummer confluent hypergeometric functions, respectively.

\end{document}